\documentclass[letter, 10pt, conference]{IEEEtran}

\IEEEoverridecommandlockouts                              

\usepackage{graphics} 
\usepackage{epsfig} 
\usepackage{mathptmx} 
\usepackage{times} 
\usepackage{amsmath} 
\usepackage{amssymb}  

\usepackage{subfig}
\usepackage{booktabs}
\usepackage{multirow}
\usepackage{tikz}
\usepackage{tikzscale}
\usepackage{pgfplots}
\usepackage{gensymb}
\usepgfplotslibrary{units}

\title{\LARGE \bf CNN-based synthesis of realistic high-resolution LiDAR data}

\author{Larissa T. Triess$^{1,2}$, David Peter$^{1}$, Christoph B. Rist$^{1}$, Markus Enzweiler$^{1}$, and J. Marius Z\"ollner$^{2,3}$%
\thanks{$^{1}$Daimler AG, Research and Development, Stuttgart, Germany}%
\thanks{$^{2}$Karlsruhe Institut of Technology, Karlsruhe, Germany}%
\thanks{$^{3}$FZI Research Center for Information Technology, Karlsruhe, Germany}%
\thanks{Primary contact: {\tt\small larissa.triess@daimler.com}}
}

\newcommand{\LossAlpha}[1]{$\mathcal{L}^\text{#1}$}
\newcommand{\LossDist}[1]{$\mathcal{L}^\text{#1}_\text{dist}$}

\newcommand{\LossFeatBlock}[1]{$\mathcal{L}_{\text{feat},#1}$}
\newcommand{\LossSC}{$\mathcal{L}_{\text{SC}}$}
\newcommand{\FM}[1]{\phi\!(#1)}

\renewcommand{\vec}[1]{\mathbf{#1}}

\begin{document}

\IEEEoverridecommandlockouts
\IEEEpubid{\makebox[\columnwidth]{978-1-7281-0559-8/19/\$31.00~\copyright2019 IEEE \hfill} \hspace{\columnsep}\makebox[\columnwidth]{ }}

\maketitle
\IEEEpubidadjcol
\pagestyle{empty}

\begin{abstract}
This paper presents a novel CNN-based approach for synthesizing high-resolution LiDAR point cloud data.
Our approach generates semantically and perceptually realistic results with guidance from specialized loss-functions.
First, we utilize a modified per-point loss that addresses missing LiDAR point measurements. Second, we align the quality of our generated output with real-world sensor data by applying a perceptual loss.

In large-scale experiments on real-world datasets, we evaluate both the geometric accuracy and semantic segmentation performance using our generated data vs. ground truth.
In a mean opinion score testing we further assess the perceptual quality of our generated point clouds.
Our results demonstrate a significant quantitative and qualitative improvement in both geometry and semantics over traditional non CNN-based up-sampling methods.
\end{abstract}

\section{INTRODUCTION}
\label{sec:introduction}

LiDAR scanners are a key enabler for autonomous driving.
They are required to have a very high resolution to provide detailed information on the environment and ensure a high detection performance.
Due to the unique properties of the LiDAR data structure, irregularity, and sparsity, it is not trivial to increase the point density of LiDAR scans.
Typical approaches start by accumulating scans over time or by the guidance of high resolution RGB camera images.
The former faces the difficulty of motion.
It is possible to eliminate this effect for static objects, since the ego-motion of the sensor
is usually known.
However, this does not account for the movement of passing objects, which remains a restriction of this approach.
For the second case, it is necessary to have a RGB camera in the sensor setup, which is not always guaranteed.
Furthermore, if the distance between the two sensors is quite large, it becomes more difficult to translate useful data from one sensor to the other.
To overcome these issues, we focus on single frame up-sampling with only one modality, the LiDAR sensor.

\begin{figure}[t]
    \centering
   	\parbox{.48\textwidth}{
   	\centering
			\includegraphics[width=.47\textwidth]{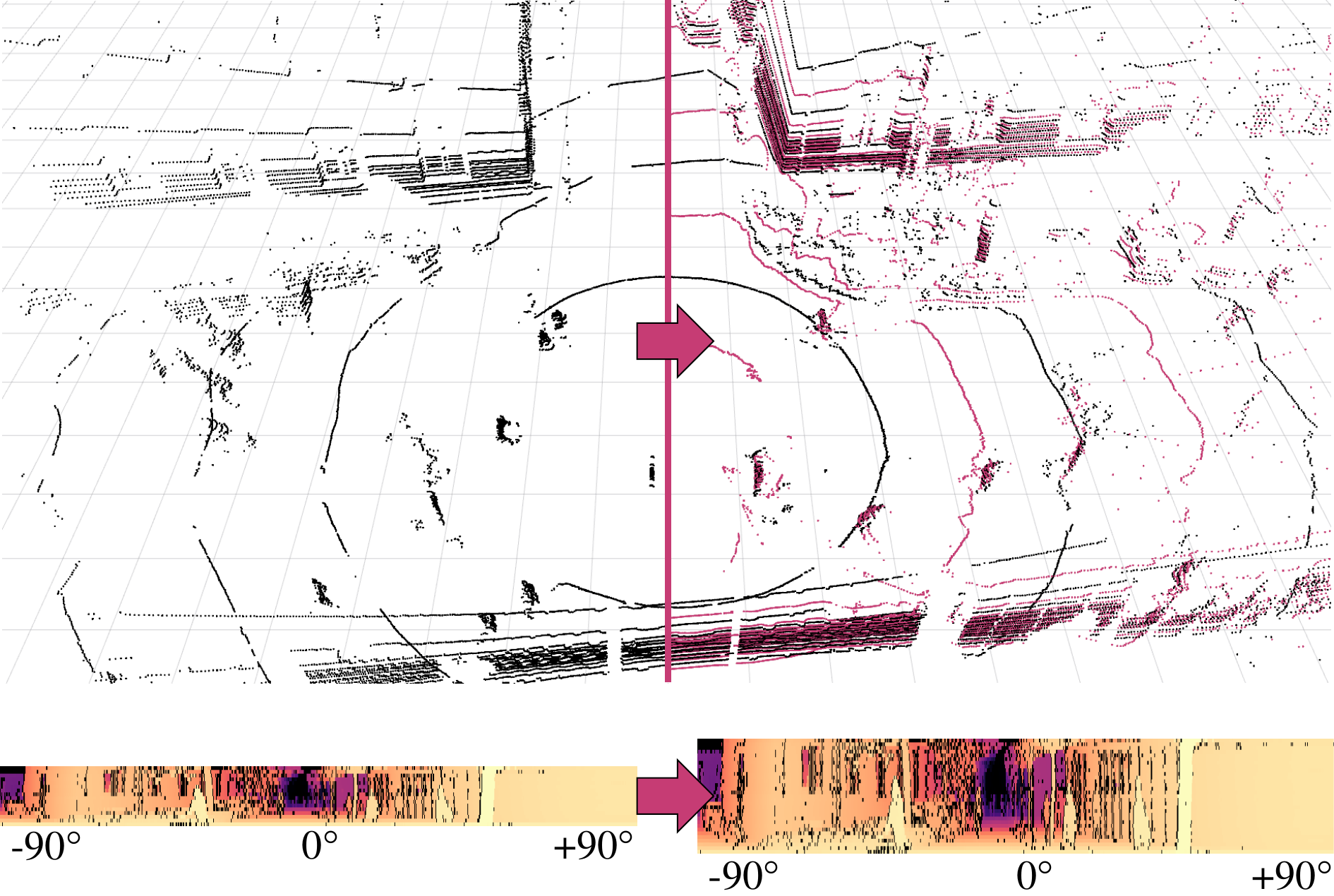}
   	}
    \caption{\textbf{Up-sampled point cloud:} The top left shows a three dimensional point cloud recorded by a Velodyne VLP32 LiDAR scanner, with every other layer removed. On the bottom left, the corresponding cylindrical projection, the LiDAR distance image, is depicted in range $-90\degree<\theta<+90\degree$. The color coding depicts closer objects in brighter colors and marks missing measurements in black. For better visibility, the vertical pixel size is five times the actual size. The right side shows the same scene synthesized with our approach. It is up-sampled with a factor of two and every other layer is colored in violet in the top image. The scene shows an urban street crossing with cars, pedestrians, bicycles, buildings and trees.}
  	\label{fig:example_LiDAR_data}
\end{figure}

\begin{figure*}[t]
	\centerline{\includegraphics[scale=1]{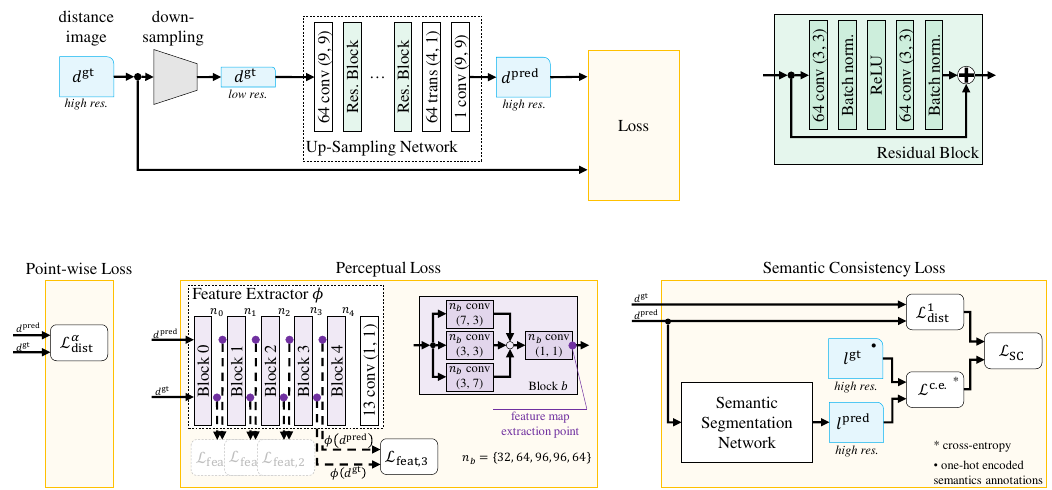}}
	\caption{\textbf{Overview on the proposed architecture:} It is divided into three separate networks. The top shows the overall architecture with a detailed view on the residual block (green). The input to the network is a down-sampled distance image of size $L/2\times W$ with information about the missing measurements. The residual up-sampling network outputs an up-sampled distance image of size $L\times W$ with in-network up-scaling. Both distance images are inputs to the loss (yellow). The bottom shows the three different loss functions under consideration (only one is used at a time).}
	\label{fig:system_overview}
\end{figure*}

LiDAR scans are generated by periodically emitting laser pulses while rotating around a vertical axis.
Detailed three dimensional geometric information about the vehicles surrounding is obtained, see fig.~\ref{fig:example_LiDAR_data}~(top).
The resulting point clouds are typically irregular and sparse in three dimensional space.
Processing high-resolution data directly with three dimensional convolutions is challenging without compromising accuracy \cite{Riegler2017}.
Given that our three-dimensional point clouds exhibit a regular structure, we can use cylindrical two-dimensional projections to represent the data in a 2D (image-like) fashion, see fig.~\ref{fig:example_LiDAR_data}~(bottom).
This allows to design two-dimensional convolutional neural networks (CNNs) which are more efficient than three-dimensional CNNs.

One observes that LiDAR sensors provide data with a high horizontal resolution, e.g. 1800~px.
However, the vertical resolution is only a small fraction of that, depending on the number of lasers within the LiDAR sensor, for example 16, 32, or 64 for commonly used Velodyne LiDARs \cite{Velodyne}.
Therefore, the objective of this work is to up-sample the vertical resolution of LiDAR scans, thereby synthesizing LiDAR data as if recorded by a scanner with more layers.
Operating in 2D preserves the regular structure in our data, see fig.~\ref{fig:example_LiDAR_data}~(bottom), which is beneficial for downstream perception algorithms.
Further, since the industry is constantly moving towards higher resolution sensors, our approach enables the re-usage of recorded (and possibly annotated) data when moving towards a higher resolution LiDAR in the future.
\newline

\section{RELATED WORK}
\label{sec:related_work}

The aim of up-sampling is to estimate the high-resolution visual output of a corresponding low-resolution input.
In this work we consider cylindrical two dimensional projections of structured LiDAR point clouds, therefore it is vital to also take into account analogous approaches on RGB images to solve this task.
A sizable amount of literature exists on RGB image up-sampling.
We focus on what we consider most relevant to this paper.
Yang et al. present a comprehensive evaluation of prevailing RGB up-sampling techniques prior to the adoption of convolutional neural networks \cite{Yang2014}.
More advanced techniques, such as SRCNN \cite{Dong2016}, outperform these traditional methods.
However, they cannot cope with data that features missing measurements, since dense input representations are required.
The traditional methods, on the other hand, can easily be applied to cylindrical LiDAR projections.
Due to their low computational complexity they can be used for real-time applications.
However, the traditional re-sampling techniques are not able to restore the high-frequency information, i.e. fine details in the resized input, due to the low-pass behavior of the interpolation filters \cite{Gavade2013}.

The literature on up-sampling three-dimensional data falls far behind the one on RGB image up-sampling.
A number of methods considers point cloud up-sampling as a depth completion task by projecting the laser scans into sparse depth maps.
They either directly operate on the depth input \cite{Uhrig2017} or require guidance, e.g. from a high-resolution camera image \cite{Dolson2010, Liu2013, Song2016, Hui2016}.
Here, the original structure of the input point cloud is lost and transformed into a high-resolution depth map at camera image resolution.

Yu et al. recently proposed PU-Net which directly operates on three dimensional point clouds \cite{Yu2018}.
The up-sampling network learns multilevel features per point and expands the point set via multi-branch convolution units.
The expanded feature is then split into a multitude of features, which are then reconstructed to an up-sampled point set.
The point set is unordered and forms a generic point cloud.
However, for our application, it is important to maintain the ordered point cloud structure provided by LiDAR sensors.
First, we are able to apply downstream perceptual algorithms which have been designed for the structured low-resolution data.
Second, it is possible to re-use valuable data recordings by up-sampling it to higher resolutions especially when new LiDAR sensors with more layers are introduced to the market. Or in the case when algorithms, like semantic segmentation \cite{Piewak2018_autolabeling} or stixels \cite{Piewak2018_stixel}, have to be adapted to the higher resolution.

The aforementioned methods all focus on optimizing a pixel level error of the prediction towards the target.
Especially in RGB images, the literature agrees that high-resolution image predictions typically do not appear visually realistic to humans \cite{Dahl2017}.
Resulting high-resolution images often lack high-frequency details and are perceptually unsatisfying in terms of failing to match the fidelity expected at the higher resolution.
Therefore, a variety of perceptual optimization methods evolved since.
Johnson et al. proposed a style transfer and up-sampling network using a perceptual loss based on VGG-16 \cite{Johnson2016, Simonyan2014}.
In contrast to previous methods, it uses an in-network re-sizing layer, making it independent from bicubic interpolation pre-processing.
In 2017, Ledig et al. proposed SRGAN which uses a perceptual loss function consisting of an adversarial loss and a content loss \cite{Ledig2017}.
The adversarial loss pushes the output to the natural image manifold using a discriminator network.
This network is trained to differentiate between super-resolved images and the origin photo-realistic images.
Additionally, a content loss was used which enforces perceptual similarity instead of similarity in the pixel space.
To the best of our knowledge, no literature on perceptual losses applied to LiDAR data exists.

Our main contributions are three-fold:
\begin{itemize}
	\item we present a CNN-based up-sampling approach that synthesizes semantically and perceptually realistic point clouds with three specialized loss functions
	\item to the best of our knowledge, our approach is first to employ perceptual losses for LiDAR based applications
	\item besides quantitative performance evaluation on large-scale real-world data, we also analyze qualitative performance through a mean opinion score study involving 30 human subjects
\end{itemize}


\section{METHOD}
\label{sec:method}

Fig.~\ref{fig:system_overview} depicts our overall system architecture in three different variants indicated by the yellow rectangles on the bottom of the figure.
The up-sampling network transforms a low resolution LiDAR scan into a corresponding high-resolution output.
This prediction is compared with the ground truth high-resolution scan in either the point-wise, perceptual or semantic consistency loss function (yellow rectangles).
An error is calculated which is then minimized by an Adam optimizer \cite{Kingma2014}.

\subsection{Up-Sampling Network}
\label{sec:method_upsampling_network}

The up-sampling network is a deep residual convolutional neural network \cite{He2016}.
It up-samples the resolution of a LiDAR distance image to produce a high-resolution output.
The output data can be understood as the equivalent of a recording from a LiDAR sensor with twice as many layers.

The design of the up-sampling network is inspired by the \textit{image transformation network} by Johnson et al. \cite{Johnson2016} and uses a fractionally strided convolution for the actual up-sampling (cf. \textit{trans} block in fig.~\ref{fig:system_overview}).
Performing the resolution change \textit{in-network} is advantageous over alternative approaches where the re-scaling is implemented in a bicubic interpolation step prior to the actual network \cite{Dong2016}, as it enables the re-scaling parameters to be learned.
In contrast to the architecture by Johnson et al., our network consists of 16 residual blocks \cite{He2016} and does not need a normalizing tanh-activation at the output layer.
Furthermore, the kernel of the fractionally strided convolution has a size of~$(4,1)$.
Following~\cite{Johnson2016}, the convolutional layers within the residual blocks are followed by spatial batch normalization and a ReLU nonlinearity.
The first and last layers use~$9\times 9$ kernels while all remaining convolutions have kernel sizes of~$3\times 3$.

The input to the network is a LiDAR scan with $L/2$ layers, represented by a two-dimensional projection of shape ${L/2\times W}$. With up-sampling factors $\left(f_i,f_j\right)=\left(2,1\right)$, the output is a high-resolution distance image of shape $L \times W$. Since the network is fully-convolutional, it can be applied to inputs of any resolution.

\subsection{Cylindrical LiDAR Projection}
\label{sec:method_depth_projection}

A LiDAR scanner determines the distance to surrounding objects by measuring the time of flight of emitted laser pulses.
The kind of scanner used in this work consists of $L$ vertically stacked send-and-receive modules which revolve around a common vertical axis.
While rotating, each module periodically measures the distance $r_{ij}$ at its current orientation which can be described by an elevation angle~$\theta_i$ and an azimuth angle~$\varphi_j$.
The indices $i=1\dots L$ and $j=1\dots W$ represent the possible discrete orientations.
Point of a full 360\degree ~rotation are referred to as frame or scan.

There are multiple ways in which a LiDAR sensor can fail to provide a point distance measurement.
First, the maximum distance is limited due to beam divergence and atmospheric absorption.
Second, outgoing lasers pulses might hit specular reflective surfaces and never return to the sensor.
Third, the laser might not be pointed towards an object at all (but towards the sky).
To account for these missing measurements, we first define the set of all \textit{valid} measurements as
\begin{equation}
\label{eq:valid_point_set}
    \mathcal{V}=\left\{(i,j) \,\middle|\, \text{reflection at $\theta_i$, $\varphi_j$ received}\right\}.
\end{equation}
A two-dimensional LiDAR distance image $d_{ij}$ can then be constructed by setting
\begin{equation}
    d_{ij} = \begin{cases}
        r_{ij} / \text{m} &(i,j)\in\mathcal{V} \\
        d^{*} &\text{otherwise}
    \end{cases}
\end{equation}
where we represent all measured ranges in units of meters~[m] and define a proxy value $d^{*}$ for the missing measurements.
The latter is necessary to provide a dense image structure for the convolutional network.
Experiments show no significant difference in choosing this value, so we set $d^{*}=0$ for simplicity, and handle it appropriately within our loss function design.
The missing measurements are one of the essential differences between RGB images and LiDAR distance images.
This prevents us from using the same methods designed for RGB image up-sampling to up-sample LiDAR distance images directly.

Note that the distance image representation is a cylindrical projection without any loss of information, as there are no mutual point occlusions.
Since all orientation angles are known, the image can always be transformed back to a 3D point cloud $\{(x_{ij},y_{ij},z_{ij}) \,|\, (i,j)\in\mathcal{V}\}$ with a spherical-to-Cartesian mapping.

\subsection{Modified Point-wise Loss}
\label{sec:method_point_loss}

In a supervised setting, up-sampling is a classic regression problem where a loss function $\smash{\mathcal{L}(\vec{d}^\text{pred}, \vec{d}^\text{gt})}$ compares the generated high-resolution distance image $\smash{\vec{d}^{\text{pred}}=\{d_{ij}^{\text{pred}}\}}$ with its corresponding ground truth counterpart $\smash{\vec{d}^{\text{gt}}=\{d_{ij}^{\text{gt}}\}}$. The most commonly used error functions for this application are the \LossAlpha{1} and \LossAlpha{2} loss functions.  In the case of LiDAR distance images, we modify these loss functions to mask the missing measurements which have been replaced by $\smash{d^{*}}$. We therefore define the modified point-wise loss functions
\begin{equation}
\label{eq:regression_loss}
	\mathcal{L}_\text{dist}^\alpha = \frac{1}{\alpha\left|\mathcal{V}\right|}\sum_{(i,j)\in\mathcal{V}}\left|d_{ij}^\text{gt}-d_{ij}^\text{pred}\right|^\alpha \quad \alpha=1,2
\end{equation}
where $\alpha=1$ describes the mean average error and $\alpha=2$ describes the mean squared error.
Refer to the leftmost loss block in fig.~\ref{fig:system_overview}.

\subsection{Perceptual Loss}
\label{sec:method_feature_loss}

The previously introduced point-wise loss encourages the network to predict high-resolution LiDAR scans where each point is close to the ground truth counterpart in a purely spatial sense.
A perfect match would be ideal in theory, but this approach can fail to output realistic point clouds in practice.
To see this, note that a perfectly realistic point cloud constructed from a slightly rotated ground truth point cloud would lead to high loss values.
Similarly, while an actual scan of a treetop looks like a seemingly random collection of points, a \LossAlpha{$\alpha$}-guided optimization will tend to produce smooth surfaces to decrease the overall distance error.
We make use of a perceptual loss function to circumvent this problem.

In order to address the shortcomings of the per-pixel losses and to allow the loss function to measure semantical and perceptual differences between LiDAR scans, the perceptual loss function utilizes a deep convolutional network itself.
This network is pre-trained for point-wise semantic segmentation in LiDAR scans \cite{Piewak2018_autolabeling}, and can therefore be used as a feature extractor which encodes semantic information.
This feature extractor~$\phi$ is used to compare the scans on a more abstract level (see middle loss block in fig.~\ref{fig:system_overview}).
To achieve semantical and perceptual similarity with the ground truth scan, the perceptual loss function propagates both scans through the feature extractor~$\phi$ and computes a $\mathcal{L}^1$ error on the resulting high-dimensional feature maps:
\begin{equation}
\label{eq:perceptual_loss}
    \mathcal{L}_\text{feat} = \sum_{c,i,j}\left|\FM{\vec{d}^\text{gt}}_{cij} - \FM{\vec{d}^\text{pred}}_{cij}\right|
\end{equation}
Here, $c$ iterates over the different channels of the feature map.
Note that the weights in the feature extractor $\phi$ stay fixed during the training.
The feature maps can be extracted at various points within the network.
In contrast to our definition of the point-wise loss function it is not necessary to exclude missing measurements from the loss calculation as the semantic feature extractor uses context information to correctly label missing input points.

\subsection{Semantic consistency loss}
\label{sec:method_semantical}

In addition to the point-wise and the perceptual losses, we propose a
\textit{semantic consistency loss function} that is designed to maintain the semantic content of the
LiDAR scan during the up-sampling process.
It uses a pre-trained semantic segmentation network (same as in~\ref{sec:method_feature_loss}) to
compare the two scans in a cross-entropy fashion (see rightmost loss block in fig.~\ref{fig:system_overview}.
To that end, it propagates the high-resolution prediction $\vec{d}^\text{pred}$ through the weight-fixed network to compute logits for the 13 semantic classes (road, person, car, building, \dots).
In $\mathcal{L}^\text{cross-entropy}$, the result is compared with the one-hot encoded ground truth annotations from the Semantics dataset.

Working with the cross-entropy loss in isolation is not enough, since the spatial structure of the
predicted point cloud is now completely unconstrained. To account for this, we compute a point-wise \LossDist{1} loss in addition (see eq.~\ref{eq:regression_loss}) and combine the two loss functions
in the following multi-task \textit{semantic consistency} (SC) loss function:
\begin{equation}
	\mathcal{L}_\text{SC}=\frac{1}{2\sigma_r}\text{\LossDist{1}}+\log\sigma_r+\frac{1}{\sigma_c}\mathcal{L}^\text{cross-entropy}+\log\sigma_c.
\end{equation}
Here, $\sigma_r$ and $\sigma_c$ are trainable variables that balance the relative weights of the
two tasks, cf.\ Kendall et al.~\cite{Kendall2017}.


\section{EXPERIMENTS}
\label{sec:experiments}

In the following, we introduce the experimental setup of our performance evaluation (section~\ref{sec:experiments_setup}) as well as a discussion of quantitative (section~\ref{sec:experiments_quantitative}) and qualitative (section~\ref{sec:experiments_qualitative}) results.

\subsection{Experimental Setup}
\label{sec:experiments_setup}

\subsubsection{Training Data}
To train our networks we use two large-scale LiDAR datasets.
The first one is the dataset introduced by Piewak et al. \cite{Piewak2018_autolabeling}, which we refer to as "Semantics" dataset.
Second, the raw dataset of the public KITTI benchmark is used ("KITTI Raw").
Details are given in Table~\ref{tab:dataset_split}.
The dataset split into training (0.62), validation (0.13), and test (0.25) has been performed on a sequence basis in order to prevent correlations between subsets.

\begin{table}[tb]
\caption{Overview on the dataset split}
\label{tab:dataset_split}
\begin{center}
\begin{tabular}{lrrr}
\toprule
 & Training & Validation & Test \\
\midrule
Semantics \cite{Piewak2018_autolabeling}	& 344,027	& 73,487	& 137,682	\\
KITTI Raw \cite{Geiger2013_KITTI}			&  28,548 	&  5,982 	&  11,499	\\
\bottomrule
\end{tabular}
\end{center}
\end{table}

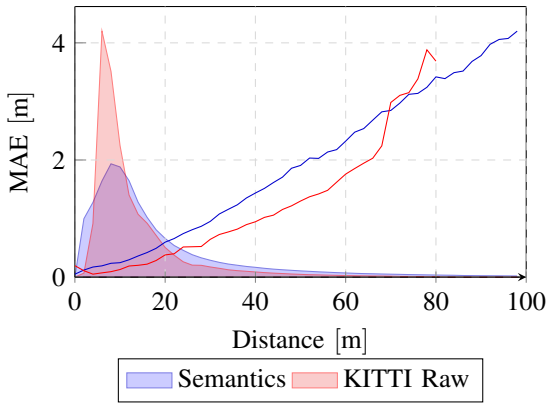
\begin{figure}[htb]
	\centering
	\parbox{.5\textwidth}{\begin{tikzpicture}
\pgfplotsset{
	width=6.0cm,
	height=3.6cm,
	compat=1.3,
	grid style={dashed,gray!30},
	scale only axis,
	xmin=0, xmax=100,
}

\begin{axis}[
	axis y line*=right,
	ymin=0,
	ticks=none,
	legend style={area legend,at={(0.5,-0.3)},anchor=north,legend columns=-1}
]

\addplot[blue!80!black,fill=blue,opacity=0.5,fill opacity=0.2] table[y=sprinkles, x=distances, col sep=comma] {data/dataset_histograms.csv};
\addplot[red,fill=red!90!black,opacity=0.5,fill opacity=0.2] table[y=kitti, x=distances, col sep=comma] {data/dataset_histograms.csv};

\legend{Semantics, KITTI Raw}

\end{axis}

\begin{axis}[
	axis y line*=left,
	axis x line=bottom,
	xlabel=Distance,
	ylabel=MAE,
	x unit=m,
	y unit=m,
	ymin=0,
	grid=both
]

\addplot[blue!80!black] table[y=sprinkles, x=distances, col sep=comma] {data/error_over_distance.csv};
\addplot[red] table[y=kitti, x=distances, col sep=comma] {data/error_over_distance.csv};

\end{axis}
\end{tikzpicture}}
	\caption{\textbf{Distance-dependent error:}
    The plot shows the mean absolute error of the \LossDist{1} network as a function of the (ground truth) distance on the Semantics and KITTI Raw datasets.
    The shaded areas depict the (normalize) distance distribution over each of the datasets.
    }
	\label{fig:histograms}
\end{figure}

Both datasets contain a variety of different scenes captured in urban, rural, and highway traffic.
However, there are important differences between the two.
Most significantly, the Semantics dataset was recorded with a Velodyne VLP32, whereas KITTI used a Velodyne HDL64 sensor.
The number in the names corresponds to the layer count (number of rows) in the LiDAR scan.
For HDL64, these layers have an equidistant spacing whereas the VLP32 has a higher layer density in the middle.
The VLP32 has a higher range, whereas the HDL64 is limited to distances lower than 80 meters.
The normalized distance distributions of both datasets are shown as shaded areas in fig.~\ref{fig:histograms}.

For the application of point cloud up-sampling it is straightforward to generate the training input and the corresponding ground truth. The presented datasets serve as our high-resolution target data with a shape of $32 \times 1800$ for Semantics and $64\times 1565$ for KITTI. In order to obtain the low resolution frames of size $L/2\times W$, every other layer is simply removed from the input scans. This procedure is different from most image up-sampling applications, where a bicubic down-sampling is used to generate the low-resolution data.
For LiDAR point clouds, this method generates unrealistic results due to the large vertical spacing between the layers.

\subsubsection{Evaluation Metrics}
We employ four different evaluation metrics for the performance assessment. The first three contribute to our quantitative results, whereas the fourth metric is based on human opinion and is used for the qualitative assessment in section~\ref{sec:experiments_qualitative}.

Since the point-wise losses \LossDist{1} and \LossDist{2} minimize the mean absolute error (MAE) and the mean squared error (MSE), respectively, we also use these two error functions as evaluation metrics by computing the corresponding average distance error on the whole test set.
Note that the numbers obtained for the \LossDist{1} error can readily be interpreted as the average point deviation in meters.
Upon convergence we select the training state with the lowest errors on the validation set and report the performance metrics on the test set of our dataset in table~\ref{tab:eval_results}.

The third quantitative metric is constituted by a semantic segmentation network.
The pre-trained network is applied to the generated point clouds, allowing us to compute the mean intersection over union~(mIoU) with respect to the ground truth annotations. This mIoU score can then be compared to the score which is obtained when using the original ground-truth high-resolution input (56.8\%, see table~\ref{tab:eval_results}).
This gives valuable information about the semantic information contained in the generated scans.
We assume that semantics are important high-level features of real-world LiDAR data and thus an indicator for realistic LiDAR scans.
The network used for this purpose has the same configuration as the feature extractor in the perceptual loss function, but uses filter counts of $n_b=\{32,64,96,96,64\}$, respectively.

Furthermore, we conducted a \textit{mean opinion score} survey with humans who evaluated the visual quality of the generated point clouds.
The results of the study, which was conducted on the Semantics dataset, are visualized in fig.~\ref{fig:mean_opinion_score}.

\begin{table}[tb]
\caption{Test set results for different metrics}
\label{tab:eval_results}
\begin{center}
\begin{tabular}{lrrrrrr}
\toprule
\multirow{3}{*}{Networks}	& \multicolumn{3}{c}{Semantics Dataset}		&& \multicolumn{2}{c}{KITTI Raw} \\
\cline{2-4} \cline{6-7} \\
							& \multicolumn{1}{c}{MSE}	& \multicolumn{1}{c}{MAE}	& \multicolumn{1}{c}{mIoU}	&& \multicolumn{1}{c}{MSE}	& \multicolumn{1}{c}{MAE}	\\
							& \multicolumn{1}{c}{[m]}	& \multicolumn{1}{c}{[m]}	& \multicolumn{1}{c}{[\%]}	&& \multicolumn{1}{c}{[m]}	& \multicolumn{1}{c}{[m]}	\\
\midrule
Ground truth		& 0.0	& 0.00	& 56.8	&& 0.0	& 0.00			\\
\midrule
Bilinear	& 88.2		& 2.29		& 34.1		&& 11.6		& 0.81	\\
Bicubic		& 97.2		& 2.59		& 28.7		&& 13.7		& 0.95	\\
Nearest neighbor	& 147.5	& 2.83	& 28.3	&& 19.6	& 0.95	\\
\midrule
\LossDist{1}		& 20.9			& \textbf{0.68}	& 34.6	&& 2.23	& \textbf{0.21}	\\
\LossDist{2}		& \textbf{17.6}	& 0.86			& 12.5	&& \textbf{1.95}	& 0.28	\\
\midrule
\LossFeatBlock{0}	&  74.1	& 1.33	& 41.2		&& \multicolumn{1}{c}{-} & \multicolumn{1}{c}{-} \\
\LossFeatBlock{1}	& 110.4	& 3.05	& 45.0		&& \multicolumn{1}{c}{-} & \multicolumn{1}{c}{-} \\
\LossFeatBlock{2} 	& 112.1	& 2.45	& \bf{49.4} 		&& \multicolumn{1}{c}{-} & \multicolumn{1}{c}{-} \\
\LossFeatBlock{3}	&  74.1	& 1.49	& 49.1		&& \multicolumn{1}{c}{-} & \multicolumn{1}{c}{-} \\
\midrule
\LossSC	& 18.1	& 0.86	& 47.4	&& \multicolumn{1}{c}{-} & \multicolumn{1}{c}{-} \\
\bottomrule
\end{tabular}
\end{center}
\end{table}

\subsubsection{Baseline and Methods}

As a baseline, we evaluated three traditional interpolation techniques: bilinear, bicubic, and nearest neighbor, see table~\ref{tab:eval_results}.

The overall architecture combined with the rightmost loss block of fig.~\ref{fig:system_overview} illustrates the setup of our training with point-wise losses.
The results of the two experiments are shown as \LossDist{1} and \LossDist{2} in table~\ref{tab:eval_results}.

\begin{figure*}[t]
   	\parbox{\textwidth}{
   	\centering
   		\subfloat[Ground Truth]{
   			\includegraphics[width=.23\textwidth]{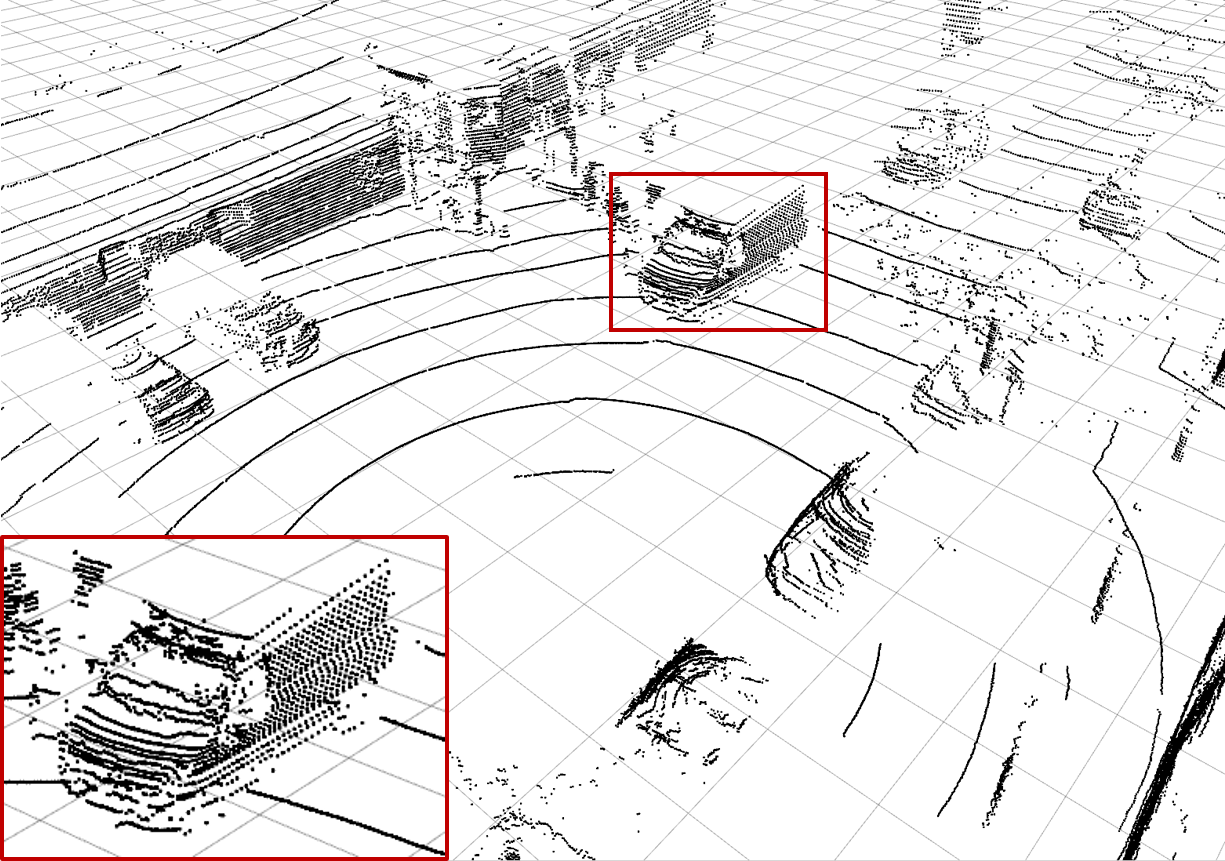}
   			\label{fig:comparison_gt}
   		}%
   		\subfloat[Low-resolution Input]{
   			\includegraphics[width=.23\textwidth]{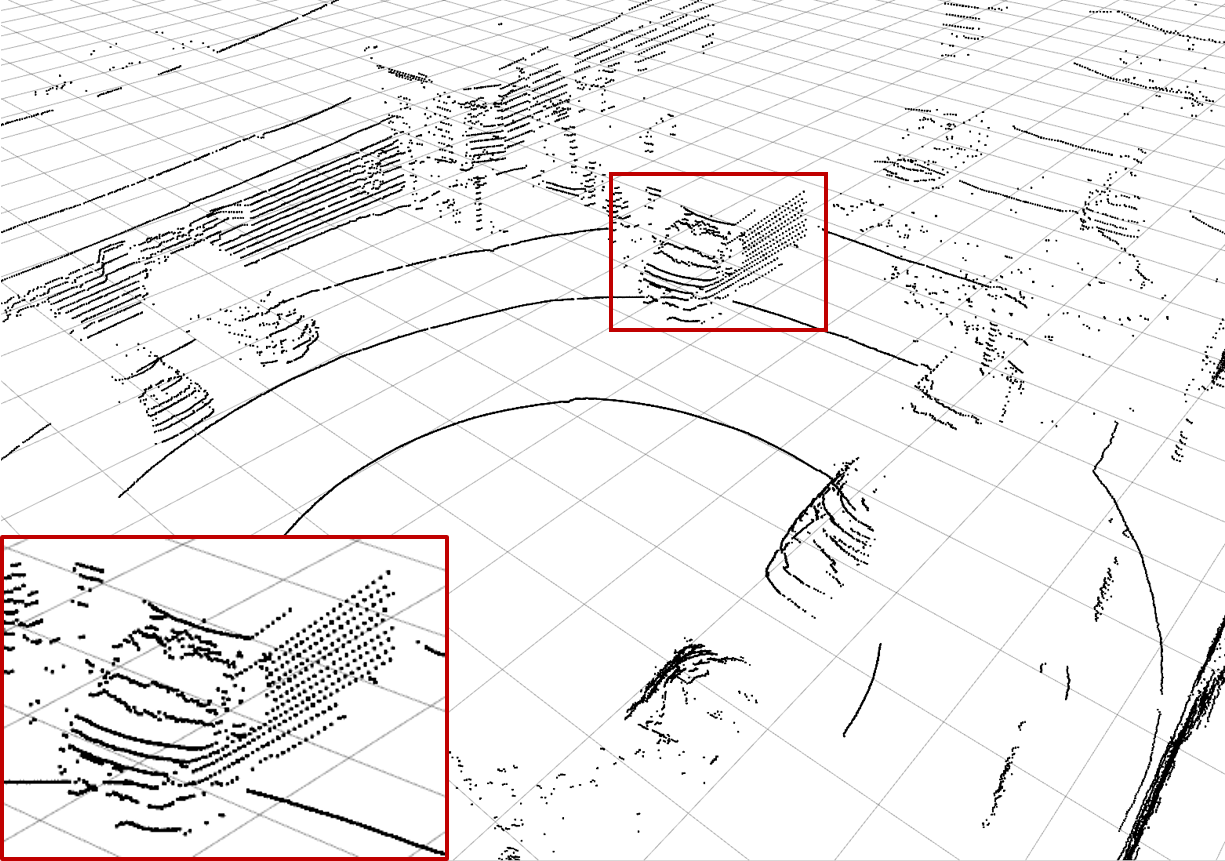}
   			\label{fig:comparison_lr}
   		}%
   		\subfloat[Bilinear Interpolation]{
   			\includegraphics[width=.23\textwidth]{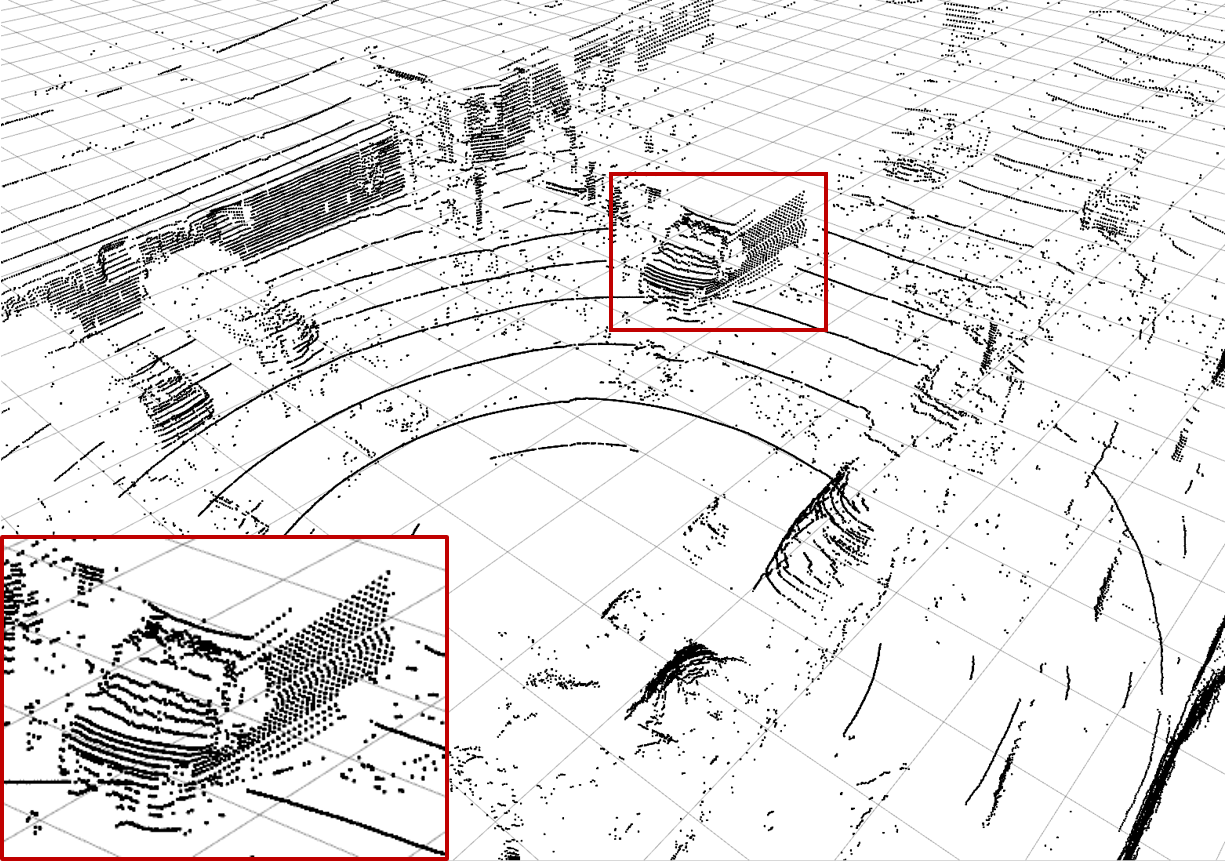}
   			\label{fig:comparison_bilinear}
   		}%
   		\subfloat[\LossDist{1} Network]{
   			\includegraphics[width=.23\textwidth]{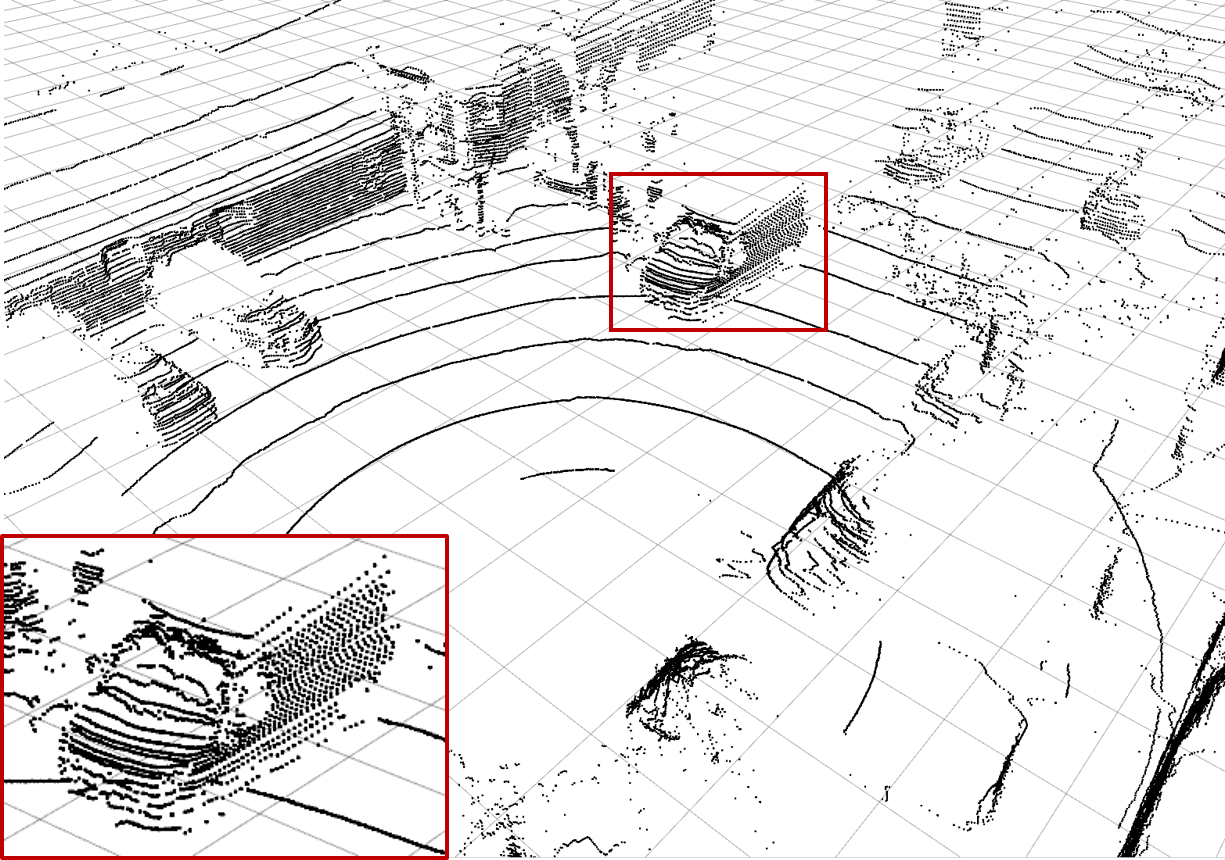}
   			\label{fig:comparison_L1}
   		}
   		
   		\subfloat[\LossDist{2} Network]{
   			\includegraphics[width=.23\textwidth]{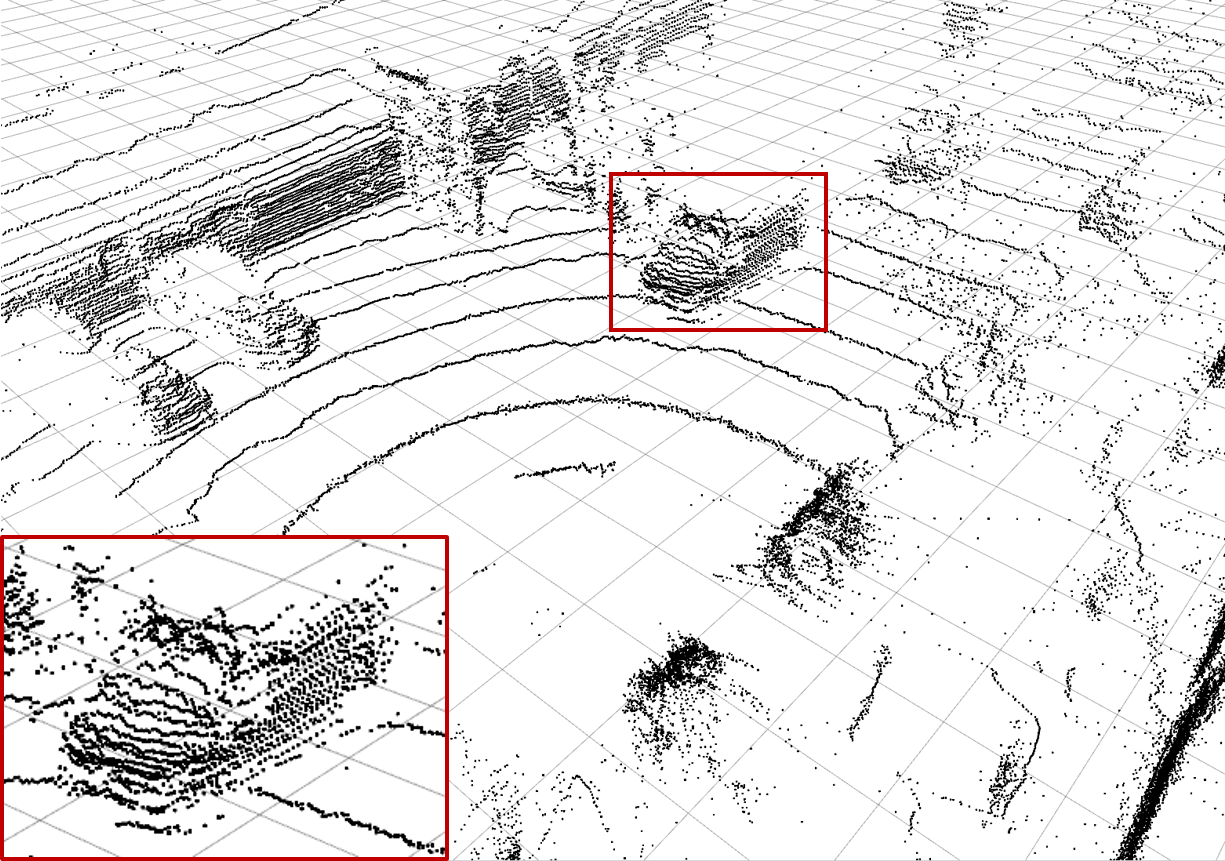}
   			\label{fig:comparison_L2}
   		}%
   		\subfloat[\LossFeatBlock{1} Network]{
   			\includegraphics[width=.23\textwidth]{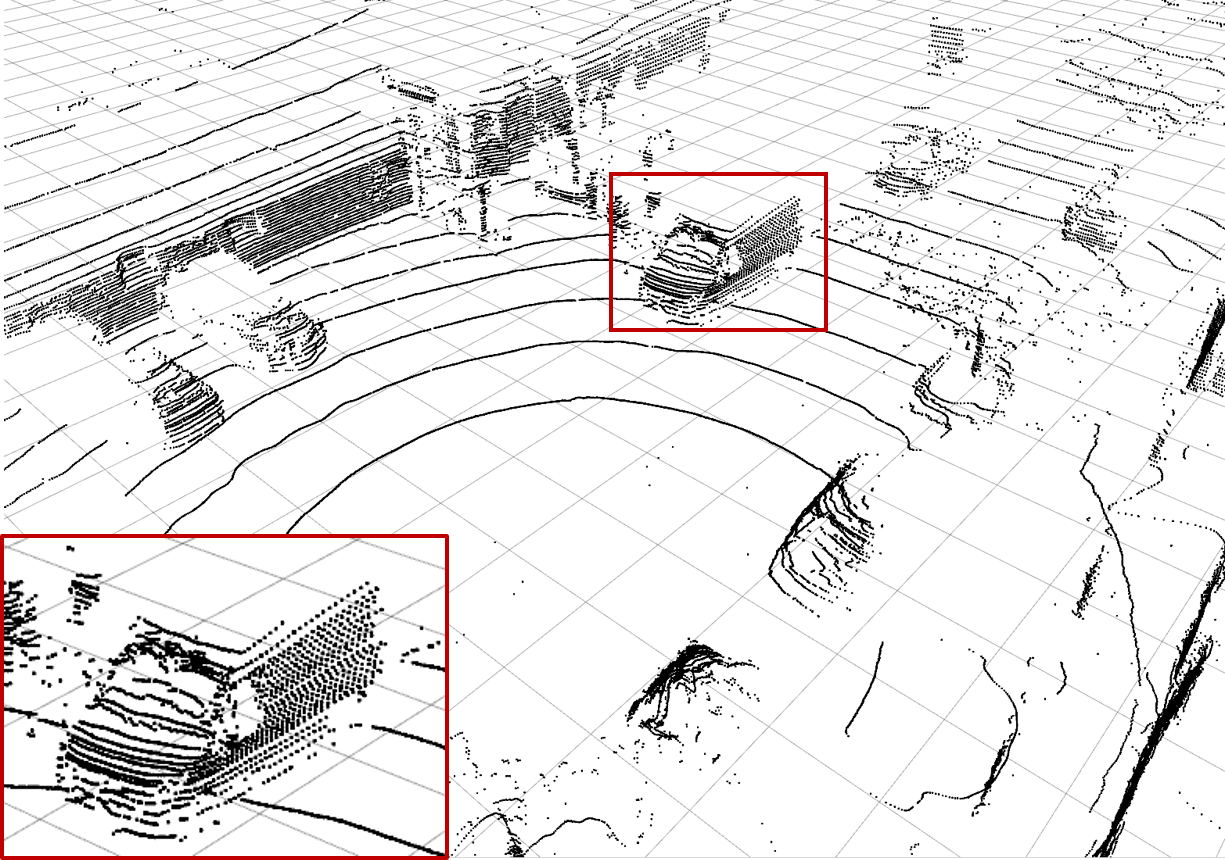}
   			\label{fig:comparison_feat1}
   		}%
   		\subfloat[\LossFeatBlock{2} Network]{
   			\includegraphics[width=.23\textwidth]{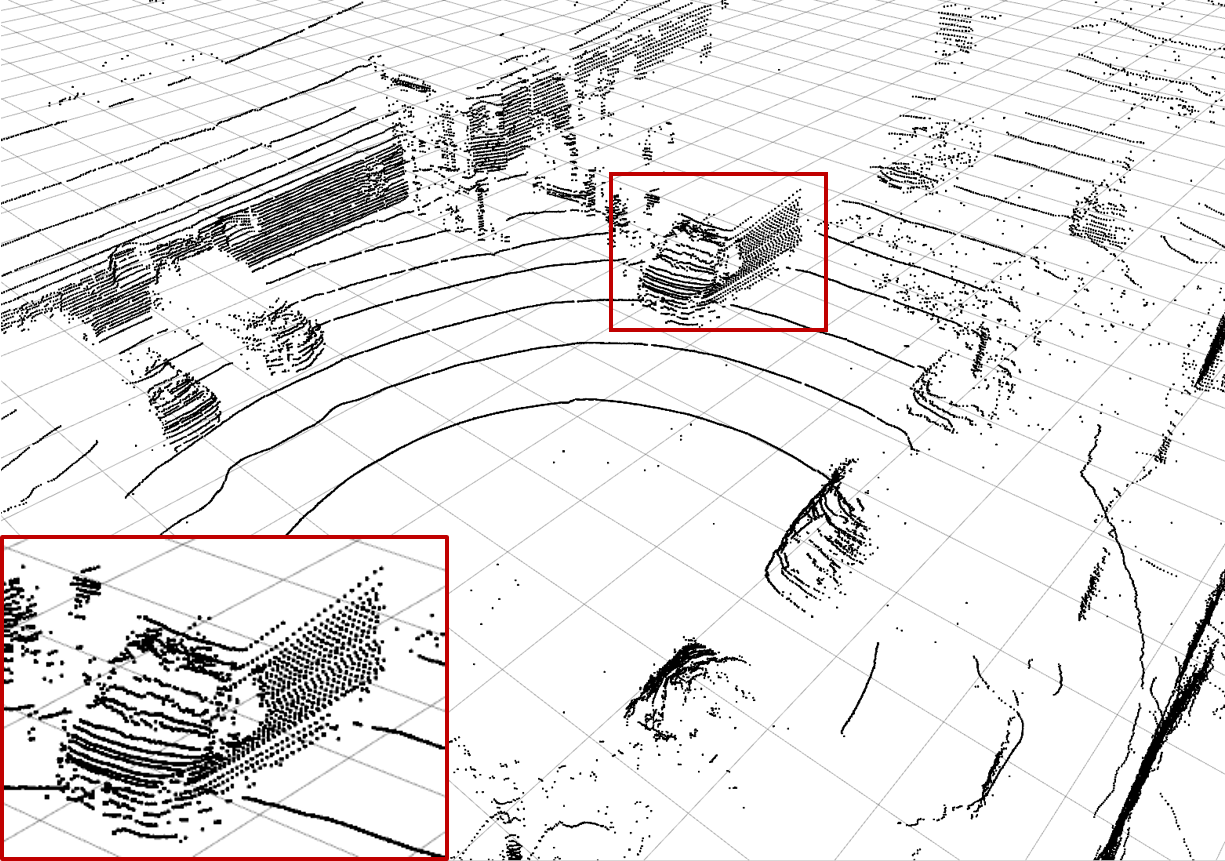}
   			\label{fig:comparison_feat2}
   		}%
   		\subfloat[Semantic consistency \LossSC]{
   			\includegraphics[width=.23\textwidth]{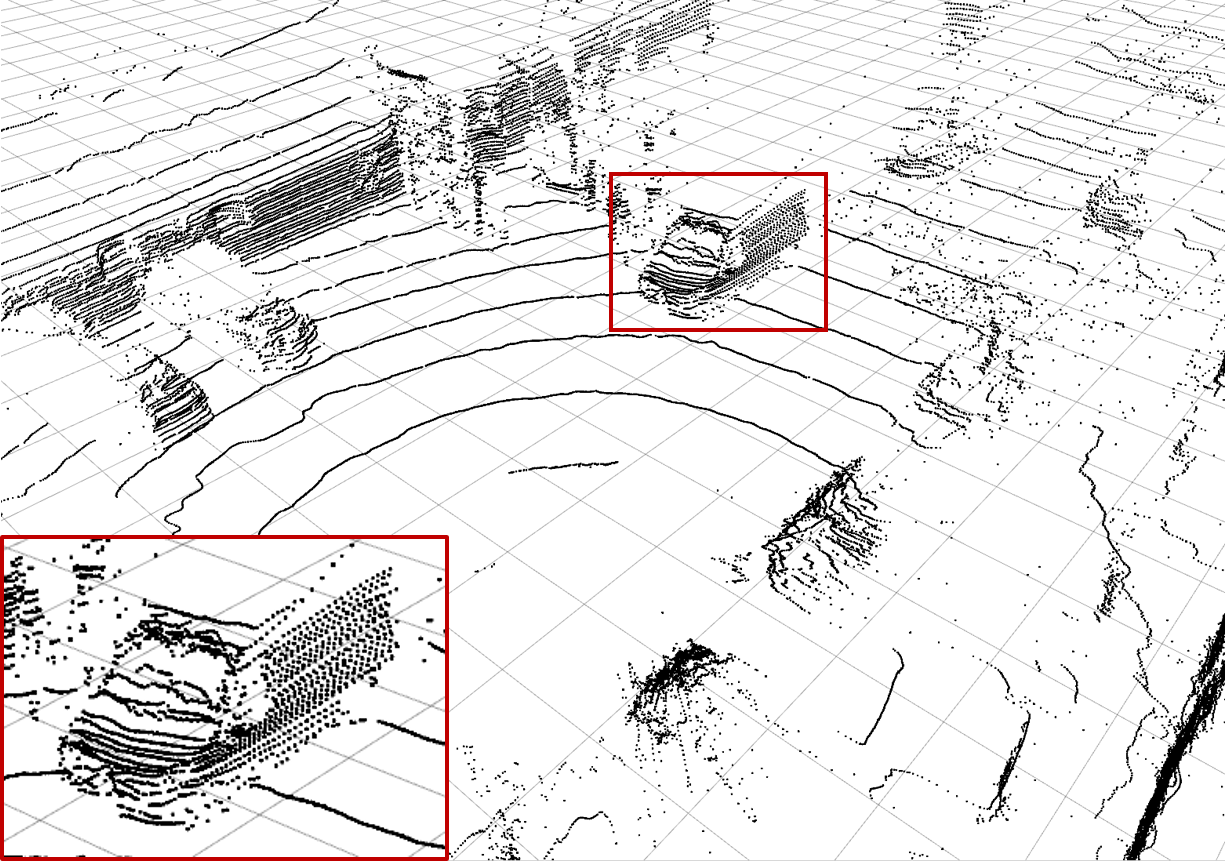}
   			\label{fig:comparison_semseg}
   		}
   	}
 	\caption{\textbf{Examples of the different methods:} Synthesize (c)~-~(h) from (b) and compare to (a). Reconstruction quality mainly differs in high frequency perturbations in object boundaries, especially \LossDist{2}~network, and overall noise level, e.g. bilinear interpolation. The red rectangle enlarges the van visible in scene.}
  	\label{fig:comparison}
\end{figure*}

The perceptual loss network has been investigated in four different variants which differ in the exact location where the feature map has been extracted.
They are illustrated by the \LossFeatBlock{b} blocks ($b=0\dots 3$) in the middle loss block of fig.~\ref{fig:system_overview}.
The same identifier is used in table~\ref{tab:eval_results}.
We encountered major up-sampling performance degradation when using large amounts of filters in the feature extractor.
This can be attributed to the storage of irrelevant information in the superfluous channels of the feature map.
We have therefore drastically reduced the amount of filters in the feature extractor to $n_b=\{32,64,96,96,64\}$, as compared to the original architecture by Piewak et. al.~\cite{Piewak2018_autolabeling}.
This change slightly reduces the performance of the semantic segmentation by 1.5 percentage points in mIoU.
The remaining difference between the original mIoU score of 60.2\% and the network used in this work (56.8\%) is due to the fact that our network is trained on distance images alone and does not use the additional reflectivity channel.
For all perceptual loss up-sampling network trainings, we use the \LossDist{1}~network weights for initialization, which speeds up the training significantly.

Finally, table~\ref{tab:eval_results} shows the results of the network trained with the semantic consistency loss \LossSC, which uses the same network architecture as the feature extractor of the perceptual trainings, in order to predict the logits for 13 semantic classes, a subset of the Cityscapes label set \cite{Cordts2016Cityscapes}.
Neither the perceptual loss networks, nor the semantic-consistency guided model can be evaluated on the KITTI dataset, due to the absence of ground truth semantic annotations for the LiDAR scans.

\subsection{Quantitative Results}
\label{sec:experiments_quantitative}
In all experiments, we encountered a significantly better performance on the KITTI dataset compared to the Semantics dataset.
We attribute this to the following reasons.
First, the KITTI dataset was recorded with a LiDAR sensor that has twice as many layers as the one used for the Semantics dataset.
This makes the interpolation easier, as neighboring points have a higher probability to lie on the same object.
Second, the HDL64 sensor used in KITTI has a smaller range. As the error generally increases with distance (see fig.~\ref{fig:histograms}), the KITTI dataset is less challenging in this respect.
Last, the VLP32 sensor does \textit{not} have an equidistant layer spacing, rendering the interpolation task on the Semantics dataset more challenging.

Considering the traditional methods first, we notice that the bilinear interpolation performs better than the bicubic interpolation.
This is in contrast to results on RGB images, where up-sampling with bicubic interpolation typically achieves better results.
The fact that this does not seem to hold for LiDAR distance images can be attributed to the low vertical resolution of the input scan, which makes next-to-nearest neighbors unlikely to contribute any useful information.
As the bilinear interpolation achieves the lowest errors, it is considered as our baseline in the following.
Unsurprisingly, it works very well for smooth surfaces but fails to properly reconstruct sharp edges.


The \LossDist{1} and \LossDist{2}-guided convolution networks are designed to address these shortcomings.
Both outperform the baseline and achieve far lower prediction errors.
Each network obtained the lowest overall error on the metric which it is designed to minimize.
Considering the mIoU, we can clearly observe a very low performance on the \LossDist{2} generated point clouds.
Taking a look at the respective label predictions reveals that a majority of the points have been labeled as vegetation or terrain.
As we will later see in the qualitative results, the \LossDist{2} generated point clouds exhibit high frequency perturbations, just like actual samples from the vegetation and terrain classes.
The reason why this effect is so prominent for the \LossDist{2} trained network lies in the mathematical definition of the mean-squared error function.
The \LossDist{2}~error penalizes larger errors more than the \LossDist{1} loss.
As the interpolation error generally increases with distance (see fig.~\ref{fig:histograms}), the optimizer tries to minimize those errors at the expense of accuracy in the near field.

In addition to those two different point-wise loss networks, we also investigated how performance changes when additionally feeding the reflectivity channel to the network.
In contrast to a semantic segmentation network, the up-sampling model did not show any performance gain.

Further, we evaluated sparse convolutions by Uhrig et al.~\cite{Uhrig2017} (with varying pooling sizes), but did not achieve competitive results on the given metrics, which can probably be explained by the fact that our input data is rather dense and only has a small fraction of missing points.

Since the perceptual networks are not specialized on minimizing the point error between prediction and target, their errors are higher than the ones for the point-wise loss networks.
However, we see a significant gain in mIoU.
The mIoU score can be improved by extracting the feature map at later stages in the network, with the highest mIoU of 49.4\% being obtained for \LossFeatBlock{2}.
The increased semantic segmentation performance leads us to the assumption that the perceptually trained networks produce perceptually more realistic point clouds than all previous methods.
This proposition will be further discussed in the qualitative assessment of the next section.

From the given experiments, we can also see that there is no method that simultaneously minimizes all three metrics.
The preferred choice of method is therefore highly dependent on the application context.

\subsection{Qualitative Results}
\label{sec:experiments_qualitative}
To gain further insight, we conducted a survey among humans to obtain a mean opinion score (MOS) for our proposed networks.
In the survey we asked LiDAR experts to evaluate the \textit{visual quality} of the generated point clouds.
In contrast to similar surveys on RGB images (showing everyday scenes) where random people have been selected \cite{Ledig2017}, the people participating in our survey were required to be familiar with LiDAR data.
This restriction was necessary because we found that laypersons could not reliably judge the point cloud quality due to a lack of domain knowledge.

\begin{figure}[tb]
	\centering
   	\parbox{.45\textwidth}{
		\includegraphics[width=.45\textwidth]{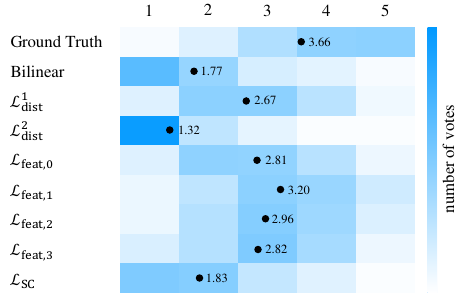}
   	}
    \caption{\textbf{Mean opinion score survey:} Color-coded distribution of mean opinion scores for ten randomly selected frames from the validation dataset. 300 votes (10 images x 30 subjects) were assessed for each method. The circular marker shows the mean opinion score (MOS).}
   	\label{fig:mean_opinion_score}
\end{figure}

The provided MOS results are illustrated in fig.~\ref{fig:mean_opinion_score}, an example scene of a selection of methods is shown in fig.~\ref{fig:comparison}. Specifically, we asked 30 subjects to assign a score from one (bad quality) to five (excellent quality) to the generated high-resolution LiDAR scans.
To this end, we rendered a view of the 3D point cloud, similar to the images in fig.~\ref{fig:comparison}.
The experts rated nine versions of each image: bilinear interpolation, networks trained with the point-wise losses \LossDist{$\alpha$} ($\alpha=1,2$), four versions of the perceptual loss \LossFeatBlock{b} ($b=0,1,2,3$), the semantic consistency loss network, and the original high-resolution LiDAR scan (ground truth).
Each subject thus rated 90 instances (nine versions of ten scenes) that were presented in a randomized fashion.
Prior to the testing, subjects were given examples for category five (ground truth) and category one (nearest neighbor interpolation, random interpolation).

Fig.~\ref{fig:mean_opinion_score} shows that the perceptually trained networks achieve higher mean opinion scores than all other networks.
The semantic-consistency network did not perform well in contrast to what the quantitative results showed.
Taking a look at the example scene in fig.~\ref{fig:comparison_semseg} shows that the generated point cloud is rather noisy.
Only the samples generated by the \LossDist{2} network, fig.~\ref{fig:comparison_L2}, exhibit even stronger high frequency perturbations, which is expressed in low MOS and mIoU scores.
The \LossDist{1} trained network achieved far better results than the bilinear interpolation or the semantic-consistency network, but still lacks the ability to fool someone to be a real sensor recording.

\begin{figure}[t]
   	\parbox{.45\textwidth}{
   	\centering
   		\subfloat[Original scan with 64 layers]{
   			\includegraphics[width=.23\textwidth]{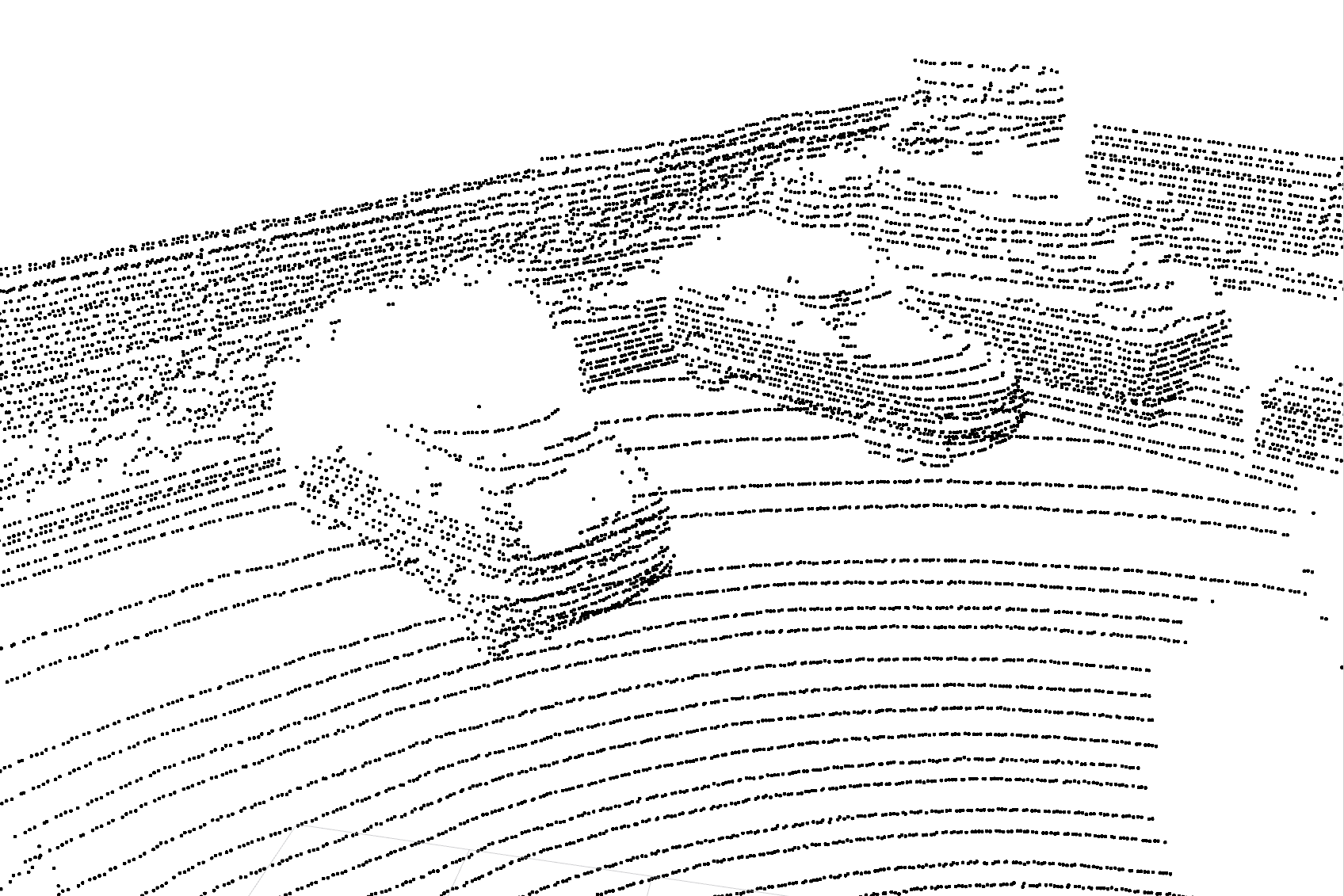}
   		}%
   		\subfloat[Up-sampled scan with 128 layers]{
   			\includegraphics[width=.23\textwidth]{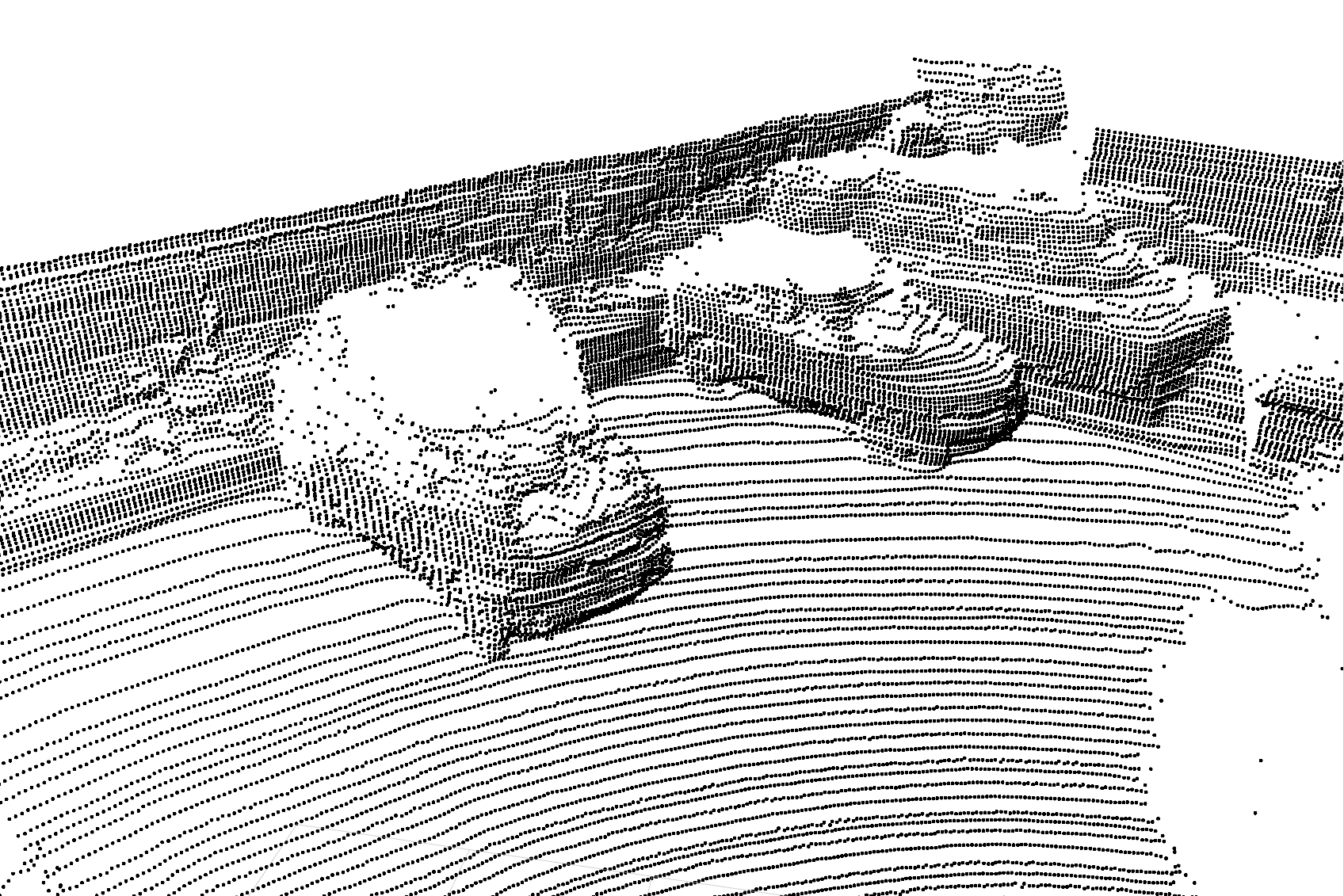}
   	}}
 	\caption{\textbf{Example scene KITTI:} The left image shows an original sensor recording of HDL64 from the KITTI Raw dataset. The right is the same scene, up-sampled to 128 layers with the \LossDist{1}~network. Since the network is fully convolutional, it is able to up-sample from 64 to 128 layers, though it was trained on 32 layers.}
  	\label{fig:example_KITTI}
\end{figure}

On average, our subjects assigned the highest ratings to the ground truth.
The rather small difference between ground truth and our \LossFeatBlock{1} approach indicates that it was sometimes difficult to distinguish between ground truth and synthesized data.
This is exactly what we wanted to achieve with the proposed approach: Synthesizing data that is almost indistinguishable from ground truth without minimizing the error on a point level but rather on a perceptual level.
We assume that our synthesis performance scales with the initial resolution of our input data.
Synthesizing 128-layer LiDAR data from 64-layer (KITTI) data as input, for example, would generate even more convincing results, see fig.~\ref{fig:example_KITTI}.
Given the lack of ground truth data, this cannot be quantitatively evaluated today and is left for future work.

\section{CONCLUSIONS}
\label{sec:conclusion}

This paper presented a novel approach for synthesizing high-resolution LiDAR scans with a high semantical and perceptual realism involving different variants of CNNs.
In extensive experiments we demonstrated that all of our system variants outperform several baseline approaches.
From a quantitative perspective, the choice of best performing model variant is highly application-specific as different models excel in different evaluation metrics with respect to geometric and semantical accuracy.
In our qualitative performance assessment, human subjects have favored model variants involving perceptual loss based on visual realism as a performance criterion.
Designing a single method that optimizes all our performance metrics at the same time is left for future work.

\addtolength{\textheight}{-10cm}   

\section*{ACKNOWLEDGMENT}

The authors thank Rainer Ott (University of Stuttgart) for valuable discussions and feedback. They also thank all participants of the mean opinion score survey.

\bibliographystyle{IEEEtran}
\bibliography{IEEEabrv,refs.bib}

\begin{thebibliography}{10}
\providecommand{\url}[1]{#1}
\csname url@rmstyle\endcsname
\providecommand{\newblock}{\relax}
\providecommand{\bibinfo}[2]{#2}
\providecommand\BIBentrySTDinterwordspacing{\spaceskip=0pt\relax}
\providecommand\BIBentryALTinterwordstretchfactor{4}
\providecommand\BIBentryALTinterwordspacing{\spaceskip=\fontdimen2\font plus
\BIBentryALTinterwordstretchfactor\fontdimen3\font minus
  \fontdimen4\font\relax}
\providecommand\BIBforeignlanguage[2]{{%
\expandafter\ifx\csname l@#1\endcsname\relax
\typeout{** WARNING: IEEEtran.bst: No hyphenation pattern has been}%
\typeout{** loaded for the language `#1'. Using the pattern for}%
\typeout{** the default language instead.}%
\else
\language=\csname l@#1\endcsname
\fi
#2}}

\bibitem{Riegler2017}
G.~Riegler, A.~O. Ulusoy, and A.~Geiger, ``Octnet: Learning deep 3d
  representations at high resolutions,'' in \emph{Proceedings of the IEEE
  Conference on Computer Vision and Pattern Recognition}, 2017.

\bibitem{Velodyne}
\BIBentryALTinterwordspacing
Velodyne. {Velodyne LiDAR}. [Online]. Available:
  \url{{https://velodynelidar.com/}}
\BIBentrySTDinterwordspacing

\bibitem{Yang2014}
C.-Y. Yang, C.~Ma, and M.-H. Yang, ``Single-image super-resolution: A
  benchmark,'' in \emph{Computer Vision -- ECCV 2014}, D.~Fleet, T.~Pajdla,
  B.~Schiele, and T.~Tuytelaars, Eds.\hskip 1em plus 0.5em minus 0.4em\relax
  Springer International Publishing, 2014, pp. 372--386.

\bibitem{Dong2016}
C.~Dong, C.~C. Loy, K.~He, and X.~Tang, ``Image super-resolution using deep
  convolutional networks,'' \emph{IEEE Transactions on Pattern Analysis and
  Machine Intelligence}, vol.~38, no.~2, pp. 295--307, Feb 2016.

\bibitem{Gavade2013}
A.~Gavade and P.~Sane, ``Super resolution image reconstruction by using bicubic
  interpolation,'' in \emph{National Conference on Advanced Technologies in
  Electrical and Electronic Systems}, 10 2014.

\bibitem{Uhrig2017}
J.~Uhrig, N.~Schneider, L.~Schneider, U.~Franke, T.~Brox, and A.~Geiger,
  ``Sparsity invariant cnns,'' in \emph{2017 International Conference on 3D
  Vision (3DV)}, Oct 2017, pp. 11--20.

\bibitem{Dolson2010}
J.~Dolson, J.~Baek, C.~Plagemann, and S.~Thrun, ``Upsampling range data in
  dynamic environments.'' in \emph{CVPR}.\hskip 1em plus 0.5em minus
  0.4em\relax IEEE Computer Society, 2010, pp. 1141--1148.

\bibitem{Liu2013}
M.~Liu, O.~Tuzel, and Y.~Taguchi, ``Joint geodesic upsampling of depth
  images,'' in \emph{2013 IEEE Conference on Computer Vision and Pattern
  Recognition}, June 2013, pp. 169--176.

\bibitem{Song2016}
X.~Song, Y.~Dai, and X.~Qin, ``Deep depth super-resolution: Learning depth
  super-resolution using deep convolutional neural network,'' in \emph{Computer
  Vision - {ACCV} 2016 - 13th Asian Conference on Computer Vision, Taipei,
  Taiwan, November 20-24, 2016, Revised Selected Papers, Part {IV}}, 2016, pp.
  360--376.

\bibitem{Hui2016}
T.-W. Hui, C.~C. Loy, and X.~Tang, ``Depth map super-resolution by deep
  multi-scale guidance,'' in \emph{Proceedings of European Conference on
  Computer Vision (ECCV)}, 2016.

\bibitem{Yu2018}
L.~Yu, X.~Li, C.-W. Fu, D.~Cohen-Or, and P.-A. Heng, ``{PU-Net}: Point cloud
  upsampling network,'' in \emph{Proceedings of IEEE Conference on Computer
  Vision and Pattern Recognition (CVPR)}, 2018.

\bibitem{Piewak2018_autolabeling}
F.~Piewak, P.~Pinggera, M.~Sch{\"{a}}fer, D.~Peter, B.~Schwarz, N.~Schneider,
  M.~Enzweiler, D.~Pfeiffer, and J.~M. Z{\"{o}}llner, ``Boosting lidar-based
  semantic labeling by cross-modal training data generation,'' in
  \emph{Computer Vision - {ECCV} 2018 Workshops - Munich, Germany, September
  8-14, 2018, Proceedings, Part {VI}}, 2018, pp. 497--513.

\bibitem{Piewak2018_stixel}
F.~Piewak, P.~Pinggera, M.~Enzweiler, D.~Pfeiffer, and M.~Z{\"{o}}llner,
  ``{Improved Semantic Stixels via Multimodal Sensor Fusion},'' in \emph{GCPR},
  2018.

\bibitem{Dahl2017}
R.~Dahl, M.~Norouzi, and J.~Shlens, ``Pixel recursive super resolution,'' in
  \emph{2017 IEEE International Conference on Computer Vision (ICCV)}, Oct
  2017, pp. 5449--5458.

\bibitem{Johnson2016}
J.~Johnson, A.~Alahi, and L.~Fei-Fei, ``Perceptual losses for real-time style
  transfer and super-resolution,'' in \emph{European Conference on Computer
  Vision}, 2016.

\bibitem{Simonyan2014}
\BIBentryALTinterwordspacing
K.~Simonyan and A.~Zisserman, ``Very deep convolutional networks for
  large-scale image recognition,'' \emph{CoRR}, vol. abs/1409.1556, 2014.
  [Online]. Available: \url{http://arxiv.org/abs/1409.1556}
\BIBentrySTDinterwordspacing

\bibitem{Ledig2017}
C.~Ledig, L.~Theis, F.~Huszar, J.~Caballero, A.~Cunningham, A.~Acosta,
  A.~Aitken, A.~Tejani, J.~Totz, Z.~Wang, and W.~Shi, ``Photo-realistic single
  image super-resolution using a generative adversarial network,'' in
  \emph{Proceedings of the IEEE Conference on Computer Vision and Pattern
  Recognition}, 2017, pp. 4681--4690.

\bibitem{Kingma2014}
D.~Kingma and J.~Ba, ``Adam: A method for stochastic optimization,''
  \emph{International Conference on Learning Representations}, 12 2014.

\bibitem{He2016}
K.~He, X.~Zhang, S.~Ren, and J.~Sun, ``Deep residual learning for image
  recognition,'' in \emph{2016 IEEE Conference on Computer Vision and Pattern
  Recognition (CVPR)}, June 2016, pp. 770--778.

\bibitem{Kendall2017}
A.~Kendall, Y.~Gal, and R.~Cipolla, ``Multi-task learning using uncertainty to
  weigh losses for scene geometry and semantics,'' in \emph{Proceedings of the
  IEEE Conference on Computer Vision and Pattern Recognition ({CVPR})}, 2018.

\bibitem{Geiger2013_KITTI}
A.~Geiger, P.~Lenz, C.~Stiller, and R.~Urtasun, ``Vision meets robotics: The
  {KITTI} dataset,'' \emph{International Journal of Robotics Research (IJRR)},
  2013.

\bibitem{Cordts2016Cityscapes}
M.~Cordts, M.~Omran, S.~Ramos, T.~Rehfeld, M.~Enzweiler, R.~Benenson,
  U.~Franke, S.~Roth, and B.~Schiele, ``The cityscapes dataset for semantic
  urban scene understanding,'' in \emph{Proc. of the IEEE Conference on
  Computer Vision and Pattern Recognition (CVPR)}, 2016.

\end{thebibliography}

\end{document}